\documentclass[twoside,twocolumn,english,aps,showpacs,prl,superscriptaddress]{revtex4-1}
\usepackage[T1]{fontenc}
\usepackage[latin9]{inputenc}
\usepackage{color}
\usepackage{bm}
\usepackage{amsmath}
\usepackage{amssymb}
\usepackage{graphicx}
\usepackage{esint}

\makeatletter

\newcommand*\LyXThinSpace{\,\hspace{0pt}}

\usepackage{times}

\@ifundefined{textcolor}{}{%
 \definecolor{BLACK}{gray}{0}
 \definecolor{WHITE}{gray}{1}
 \definecolor{RED}{rgb}{1,0,0}
 \definecolor{GREEN}{rgb}{0,1,0}
 \definecolor{BLUE}{rgb}{0,0,1}
 \definecolor{CYAN}{cmyk}{1,0,0,0}
 \definecolor{MAGENTA}{cmyk}{0,1,0,0}
 \definecolor{YELLOW}{cmyk}{0,0,1,0}
}

\makeatother

\usepackage{babel}
\begin{document}

\title{Circular Phonon Dichroism in Weyl Semimetals}

\author{Donghao Liu}

\affiliation{International Center for Quantum Materials, School of Physics, Peking
University, Beijing 100871, China}

\author{Junren Shi}
\email{junrenshi@pku.edu.cn}

\affiliation{International Center for Quantum Materials, School of Physics, Peking
University, Beijing 100871, China}

\affiliation{Collaborative Innovation Center of Quantum Matter, Beijing 100871,
China}
\begin{abstract}
We derive the phonon dynamics of magnetic metals in the presence of
strong spin-orbit coupling. We show that both a dissipationless viscosity
and a dissipative viscosity arise in the dynamics. While the dissipationless
viscosity splits the dispersion of left-handed and right-handed circularly
polarized phonons, the dissipative viscosity damps them differently,
inducing circular phonon dichroism. \textcolor{black}{The effect offers
a new degree of manipulation of phonons, }\textcolor{black}{\emph{i.e.}}\textcolor{black}{,
the control of the phonon polarization.} We investigate the effect
in Weyl semimetals. We find that there exists strong circular phonon
dichroism in Weyl semimetals breaking both the time-reversal and the
inversion symmetry, making them potential materials for realizing
the acoustic circular polarizer. 
\end{abstract}


\date{\today}

\maketitle
\noindent \textit{\textcolor{black}{Introduction}}\textcolor{black}{.\textemdash Phonons
are quasiparticles of the lattice vibrations, responsible for transmission
of heat and sound. In recent years, phononics, which aims to control
and manipulate phonons, emerges as a new research field~\cite{phononics-1,phononics-2}.
Prototypes of phononic devices such as acoustic diode have been proposed
and built~\cite{Acoustic diode theory,acoustic rectifier,Sonic-Crystal-Based Acoustic Diode Theory,Bifurcation-based acoustic rectification Experiment}.
Further, discoveries of topological materials~\cite{Hasan,Qi,Jia2016}
offer the new inspiration of topological manipulation of phonons~\cite{Topological Nature of the Phonon Hall Effect},
opening new avenues for the design of phononic devices based on topological
phonon systems~\cite{mechanical acoustic topological,acoustic topological insulator,Sound Isolation and Giant Linear Nonreciprocity,topological acoustics,circulating air flow,graphene-like resonator lattice,topological acoustics non-fluid,Synthetic gauge flux}.
To achieve a degree of control of phonons on par with that of electrons
in electronics, it is essential to find new ways of the phonon manipulation.
One of notable progresses is the discovery of the phonon Hall effect
(PHE)~\cite{Strohm,Inyushkin,Topological Nature of the Phonon Hall Effect,Qin},
which demonstrates a novel possibility of controlling phonons directly
by a magnetic field, even though phonons are charge-neutral quasi-particles.
The possibility is inspiring because compared to the conventional
mechanic ways of controlling phonons such as the circularly moving
background~\cite{Sound Isolation and Giant Linear Nonreciprocity,topological acoustics,circulating air flow,graphene-like resonator lattice},
an electromagnetic field is clearly advantageous. Theoretically, it
is found that the coupling between phonons and the magnetic field
is induced by the dissipationless viscosity~\cite{Qin,Hoyos,Fradkin,Hughes,Barkeshli,Shapourian},
a Berry phase effect which emerges in systems with strong spin-orbit
coupling (SOC). Unfortunately, the current theoretical understanding,
which is based on the adiabatic assumption and the Born-Oppenheimer
approximation~\cite{Qin}, is valid only for insulators. Since metals
are ubiquitous and preferred in many situations, it is highly desirable
to develop a theory for metals as well.}

\textcolor{black}{In this Letter, we derive the general phonon dynamics
of metals in the presence of strong SOC. We show that besides the
dissipationless viscosity, a dissipative viscosity also arises in
the phonon dynamics of metals. While the dissipationless viscosity
splits the dispersion of left-handed and right-handed circularly polarized
(LCP and RCP) phonons~\cite{L Zhang1,L Zhang2,helicity-resolved Raman},
the dissipative viscosity will damp them differently, inducing circular
phonon dichroism. The effect offers a new degree of manipulation of
phonons, }\textcolor{black}{\emph{i.e.}}\textcolor{black}{, the control
of the phonon polarization. We apply the theory to Weyl semimetals
(WSMs). We show that a Weyl node in a WSM selectively absorbs LCP
or RCP phonons, depending on its chirality. However, in a WSM breaking
time-reversal symmetry ($\mathcal{T}$) but preserving inversion symmetry
($\mathcal{I}$), the total effect of a pair of the Weyl nodes of
opposite chirality is largely cancelled. We show that further breaking
$\mathcal{I}$ will unearth the giant circular phonon dichroism inherent
to each of the Weyl nodes, greatly enhancing its effect. It makes
them potential materials for realizing the acoustic circular polarizer
that converts an injected acoustic wave into a circularly polarized
acoustic wave.}

\noindent \textit{\textcolor{black}{Phonon dynamics}}\textcolor{black}{.\textemdash }We
start by deriving the general lattice dynamics of magnetic metals.
We treat a metal as a collection of electrons and ions, and the coupling
between the electrons and the ions is described by an electron-ion
interacting potential $v_{ei}(\bm{r}-\bm{R})$~\cite{Mahan2000},
where $\bm{r}$ ($\bm{R}$) denotes the position of an electron (ion).
The collection of the ion positions can be decomposed into $\left\{ \boldsymbol{R}\right\} \equiv\left\{ \boldsymbol{R}_{n}^{0}+\boldsymbol{u}_{n},n=1,\ldots,N\right\} $,
where $\boldsymbol{u}_{n}$ is the displacement from an equilibrium
position $\boldsymbol{R}_{n}^{0}$ and $N$ is the total number of
the ions. For simplicity and without loss of generality, we consider
a monatomic lattice with an atomic mass $M$. The motion of the ions
can be treated classically. The equation of motion reads: 
\begin{equation}
M\ddot{u}_{n}^{\alpha}=-\sum_{n'\beta}\Phi_{n\alpha n'\beta}u_{n'}^{\beta}+\left\langle \hat{T}_{n,\alpha}\right\rangle ,\label{eq of motion}
\end{equation}
where $\alpha,\beta=x,y,z$, and the right hand side of the equation
is the total force acting on an atom. It contains two parts of contributions:
a direct ion-ion interaction with respect to the displacement $u_{n}^{\alpha}$
under the harmonic approximation (the first term) and a force exerted
by electrons $\langle\hat{T}_{n,\alpha}\rangle$ with $\hat{T}_{n,\alpha}\equiv-\partial\hat{V}_{ei}/\partial u_{n}^{\alpha}$,
where $\hat{V}_{ei}$ is the total electron-ion interaction energy.
The expectation value $\langle\hat{T}_{n,\alpha}\rangle$ should be
evaluated in the electron subsystem subjected to a time-dependent
ion fields $\hat{V}_{ei}\left(t\right)\approx\sum_{n\alpha}(\partial\hat{V}_{ei}/\partial u_{n}^{\alpha})_{\{\bm{u}\}=0}u_{n}^{\alpha}\left(t\right)$,
which is treated as a perturbation to the same order of the harmonic
approximation. The electron-ion force can be obtained by using the
linear response theory: 
\begin{equation}
\left\langle \hat{T}_{n,\alpha}\left(t\right)\right\rangle =-\sum_{n'}\int_{-\infty}^{t}dt'\mathcal{G}_{\alpha\beta}^{R}\left(\boldsymbol{R}_{n}-\boldsymbol{R}_{n'},t-t'\right)u_{n'}^{\beta}\left(t'\right),
\end{equation}
where 
\begin{align}
\mathcal{G}_{\alpha\beta}^{R}\left(\boldsymbol{R}_{n}-\boldsymbol{R}_{n'},t-t'\right)= & -\frac{i}{\hbar}\theta\left(t-t'\right)\nonumber \\
 & \times\left\langle \left[\hat{T}_{n,\alpha}\left(t\right),\hat{T}_{n',\beta}\left(t'\right)\right]\right\rangle _{0}\label{eq:response}
\end{align}
is a retarded response function. In the momentum and frequency domain,
the equation of motion can be written as: 
\begin{equation}
\omega^{2}u_{\boldsymbol{q}}^{\alpha}=\sum_{\beta}\left[\Phi_{\alpha\beta}\left(\boldsymbol{q}\right)+\mathcal{G}_{\alpha\beta}\left(\boldsymbol{q},\omega\right)\right]u_{\boldsymbol{q}}^{\beta},
\end{equation}
where $\Phi_{\alpha\beta}\left(\boldsymbol{q}\right)$ and $\mathcal{G}_{\alpha\beta}\left(\boldsymbol{q},\omega\right)$
are the Fourier transforms of $\Phi_{n\alpha n'\beta}/M$ and $\mathcal{G}_{\alpha\beta}^{R}\left(\boldsymbol{R}_{n},t\right)/M$,
respectively. We note that the harmonic approximation we adopt here
is sufficient for most of solids as long as they are not close to
their melting points. For systems in which the anharmonicity is non-negligible,
more novel approaches may be required~\cite{Errea2013,Borinaga2016}.

To proceed, we expand $\mathcal{G}_{\alpha\beta}\left(\boldsymbol{q},\omega\right)$
as a series in powers of $\omega$. This is because the phonon energy
$\hbar\omega$ is much smaller than the energy scale of electrons:
$\hbar\omega\ll\varepsilon_{F}$, where $\varepsilon_{F}$ is the
Fermi energy of electrons. One expects that the response function
does not change drastically in the small energy scale of $\hbar\omega$.
We expand $\mathcal{G}_{\alpha\beta}\left(\boldsymbol{q},\omega\right)$
as: 
\begin{multline}
\mathcal{G}_{\alpha\beta}\left(\boldsymbol{q},\omega\right)\approx\mathcal{G}_{\alpha\beta}^{0}\left(\boldsymbol{q}\right)+i\omega G_{\alpha\beta}\left(\boldsymbol{q}\right)-2i\omega\gamma_{\alpha\beta}\left(\boldsymbol{q},\omega\right).\label{three term}
\end{multline}
The meanings of these terms are explained as follows. The first two
terms are from the Hermitian part of $\mathcal{G}_{\alpha\beta}\left(\boldsymbol{q},\omega\right)$,
which is expanded to the first order of $\omega$. The first term
characterizes the screening effect of the electrons to the ion field.
Combining with $\Phi_{\alpha\beta}\left(\boldsymbol{q}\right)$, it
gives rise to the dynamical matrix which determines phonon dynamics
in ordinary $\mathcal{T}$ invariant systems. The second term involves
an anti-Hermitian matrix with elements $G_{\alpha\beta}(\boldsymbol{q})\equiv1/(2i)[d(\mathcal{G}_{\alpha\beta}+\mathcal{G}_{\beta\alpha}^{\ast})/d\omega]_{\omega\rightarrow0}$.
In an insulator, it becomes exactly the effective magnetic field for
phonons (or the dissipationless viscosity) defined in Ref.~\cite{Qin}
(see Supplementary Materials \cite{supplementary}). Therefore, it
is a natural generalization for the definition of the dissipationless
viscosity in a metal. Finally, the last term is derived from the anti-Hermitian
part of $\mathcal{G}_{\alpha\beta}\left(\boldsymbol{q},\omega\right)$,
and gives rise to the damping (absorption) of phonons. It exists only
in metals in which phonons can excite electron-hole pairs and be dissipated.
It is interpreted as a dissipative viscosity. The causality relates
the dissipationless viscosity and the dissipative viscosity by a Kramers-Kronig
relation~\cite{supplementary}: 
\begin{equation}
G_{\alpha\beta}(\boldsymbol{q})=2i\mathcal{P}\int_{0}^{\infty}d\nu\frac{\gamma_{\alpha\beta}\left(\boldsymbol{q},\nu\right)-\gamma_{\beta\alpha}\left(-\boldsymbol{q},\nu\right)}{\pi\nu},\label{eq:KK}
\end{equation}
where $\mathcal{P}$ denotes the Cauchy principal value. The explicit
formulas of $G_{\alpha\beta}\left(\boldsymbol{q}\right)$ and $\gamma_{\alpha\beta}\left(\boldsymbol{q},\omega\right)$
for a non-interacting system are shown in Supplementary Materials~\cite{supplementary}.

By symmetry arguments, we can show that the phonon circular dichroism
emerges in a magnetic metal with SOC. We consider the structure of
the matrix of the $\gamma$ coefficients for an acoustic wave propagating
along the magnetization axis ($z$-axis) with $\bm{q}=(0,0,q_{z})$.
If the magnetization axis has at least three-fold rotational symmetry,
we can apply a rotation in the symmetry group and transform the $\gamma$
matrix by $\gamma\rightarrow R^{-1}\gamma R$, where $R$ is the transformation
matrix of the rotation. Since the $\gamma$-matrix is invariant under
the symmetry operation, it yields that $\gamma_{xx}=\gamma_{yy}$,
$\gamma_{xy}=-\gamma_{yx}$, $\gamma_{xz}=\gamma_{yz}=\gamma_{zx}=\gamma_{zy}=0$.
Because $\gamma$ is a hermitian matrix, we must have: 
\begin{equation}
\gamma=\left[\begin{array}{ccc}
\gamma^{D} & -i\gamma^{A} & 0\\
i\gamma^{A} & \gamma^{D} & 0\\
0 & 0 & \gamma^{Z}
\end{array}\right],
\end{equation}
where $\gamma^{D}$, $\gamma^{Z}$, and $\gamma^{A}$ are real damping
coefficients. The off-diagonal elements with the anomalous damping
coefficient $\gamma^{A}$ make our system distinct from an ordinary
metal. Because the off-diagonal elements are purely imaginary, they
are present only when the system breaks $\mathcal{T}$ and has a non-vanishing
SOC. With the left-handed and right-handed circular polarization vectors
$1/\sqrt{2}(\begin{array}{ccc}
1 & i & 0\end{array})^{T}$ and $1/\sqrt{2}(\begin{array}{ccc}
1 & -i & 0\end{array})^{T}$ in the long-wave limit, we find that the damping coefficients for
LCP and RCP phonons are $\gamma^{L}=\gamma^{D}+\gamma^{A}$ and $\gamma^{R}=\gamma^{D}-\gamma^{A}$,
respectively. As a result, the phonon circular dichroism emerges.

\noindent \textit{Circular phonon dichroism}\textcolor{black}{{} }\textit{\textcolor{black}{in
WSMs}}\textcolor{black}{.\textemdash }With the general formalism established
and the concepts clarified, we apply the theory to $\mathcal{T}$
broken WSMs. A WSM is a three-dimensional crystal that hosts pairs
of band-touching Weyl nodes in the reciprocal space~\cite{Jia2016}.
For a Weyl node, we adopt an electron-phonon coupled low-energy effective
Hamiltonian similar to that introduced in Ref.~\cite{Hughes}: 
\begin{equation}
H_{\pm}=\pm\left[\frac{1}{2}v_{F}\left(p_{\mu}e_{a}^{\mu}+e_{a}^{\mu}p_{\mu}\right)\sigma^{a}-\lambda\right]-b_{3}\sigma^{3},\label{Weyl Hamiltonian}
\end{equation}
where $a,\mu=1,2,3$, the sign ``$\pm$'' denotes the chirality
of the Weyl node, $v_{F}$ is the Fermi velocity, $\sigma^{a}$ is
the Pauli matrix. $e_{a}^{\mu}=\delta_{a}^{\mu}-\partial u_{a}/\partial x_{\mu}$
is the tetrad field which describes the local stretching and rotation
of lattice structure~\cite{Fradkin,Hughes}. $\mathcal{T}$ is broken
by the term $b_{3}\sigma^{3}$, which is induced either by a spontaneous
magnetization of the system or an external magnetic field coupling
to spins. The term splits a doubly degenerate Dirac node into two
separated Weyl nodes of opposite chirality with a distance $2k_{0}=2b_{3}/\hbar v_{F}$
along the $z$ direction of the reciprocal space. $\lambda$ is a
Dresselhaus-like SOC term, which breaks $\mathcal{I}$ while preserving
the rotation symmetry~\cite{Zyuzin}. It shifts the two Weyl nodes
to different energies. We note that Eq.~(\ref{Weyl Hamiltonian})
is actually a minimum form of the coupling between electrons and the
deformation. It is parameter-free and purely geometrical. Alternatively,
one could introduce the coupling as elastic gauge fields~\cite{Cortijo},
which, as we show in the Supplementary Materials~\cite{supplementary},
will only induce minor quantitative corrections to the results presented
in the following.

The dissipationless viscosity $G_{\alpha\beta}(\bm{q})$ is a property
of the Fermi sea, and a quantitative determination of $G_{xy}\left(\boldsymbol{q}\right)$
requires knowledge of high energy details which are not present in
the low energy effective Hamiltonian Eq.~(\ref{Weyl Hamiltonian}).
A renormalization of the quantity was discussed in Refs.~\cite{Fradkin,Hughes}
by assuming that the dissipationless viscosity is non-vanishing only
when the filled band of the system has a nonzero Chern number. However,
both the calculations of tight-binding models~\cite{Barkeshli,Shapourian}
and general theoretical considerations~\cite{Qin} suggest that the
dissipationless viscosity is in general non-vanishing in magnetic
metals with SOC, even when the Chern number is zero. The nonzero $G_{xy}\left(\boldsymbol{q}\right)$
splits doubly degenerate transverse acoustic (TA) phonon modes along
high symmetry axes into an LCP branch and an RCP branch. In the long-wave
limit, one can show that for systems breaking $\mathcal{T}$ but preserving
$\mathcal{I}$, the splitting has the form of $\omega_{L/R}(\bm{q})=c_{T}q\pm\alpha qq_{z}$,
where $c_{T}$ is the speed of sound for the TA modes, and $\alpha$
is a phenomenological constant~\cite{Qin}. To the lowest order of
the wave-vector $\bm{q}$, the splitting can be ignored. On the other
hand, in systems breaking both $\mathcal{T}$ and $\mathcal{I}$,
nonzero $G_{xy}\left(\boldsymbol{q}\right)$ will induce a difference
in the speed of sound for the LCP and RCP modes.

\begin{figure}
\includegraphics[width=0.54\columnwidth]{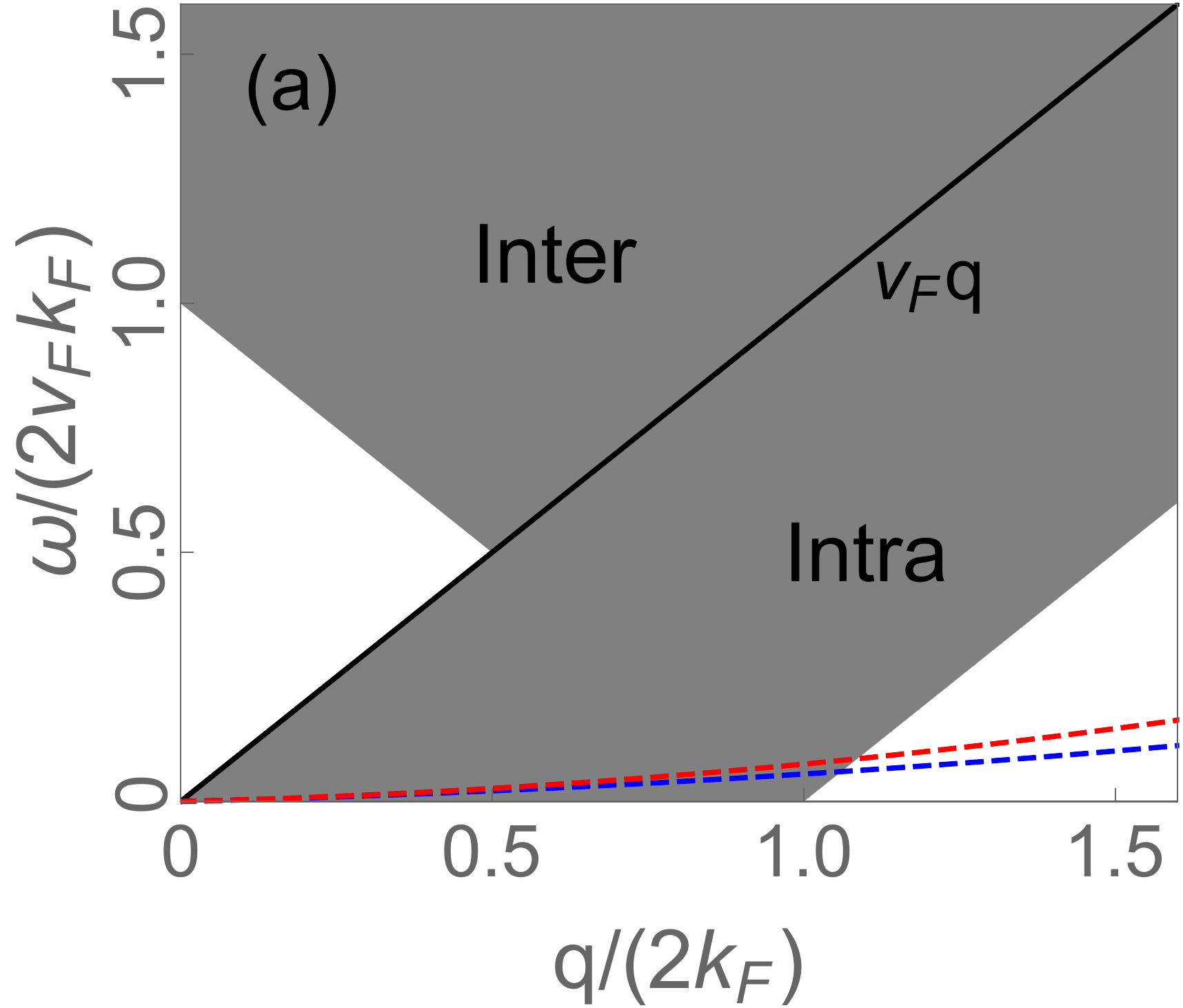}~\includegraphics[width=0.45\columnwidth]{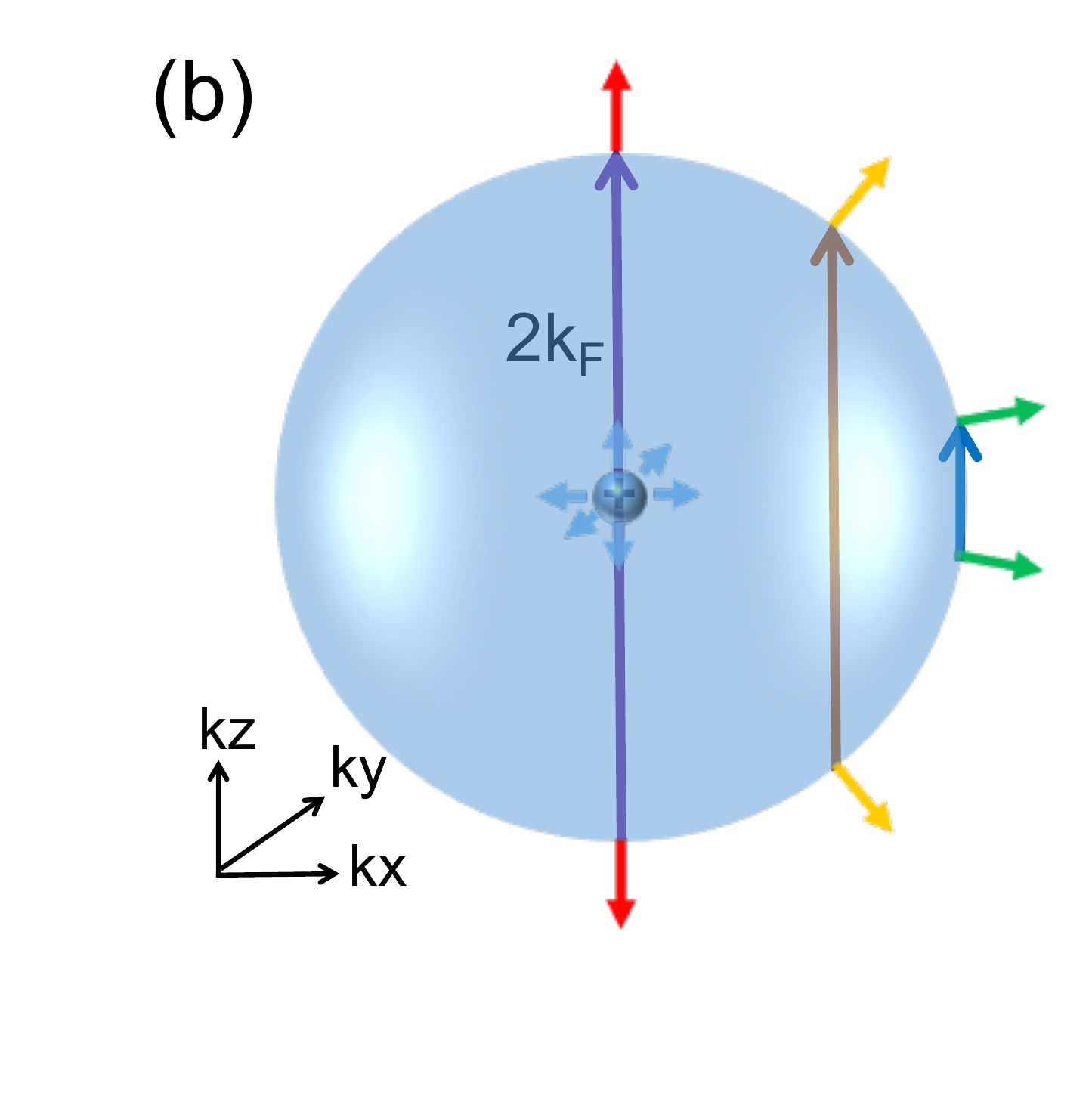}

\caption{\label{fig1} (a) Allowed energy-momentum regions of electron-hole
pair excitations for a Weyl node, indicated as shaded areas. The dashed
lines show the dispersion of the two circularly polarized phonon modes.
The ``intra'' and ``inter'' denote intra-band and inter-band pairing,
respectively. $k_{F}$ is the Fermi momentum of electrons. (b) Spin
configurations of the initial and final states in transitions induced
by phonons. We show three representative cases for different phonon
wave vectors: a small one, an intermediate one, and the largest one
allowing for the phonon absorption.}
\end{figure}

Next, we consider the damping of phonons in WSMs. We assume that the
Weyl nodes are well separated in the reciprocal space: $k_{0}\gg k_{F}$,
where $k_{F}$ is the Fermi wave number. We further limit our consideration
to long-wave phonons with $|\bm{q}|\ll2k_{0}$. As a result, all electron
transitions induced by the phonons are within one of the Weyl nodes,
and the Weyl nodes can be considered separately. The electron dispersion
near a Weyl node consists of an upper band and a lower band separated
by the band touching point. In principle, an electron absorbing a
phonon could make either an intra-band transition or an inter-band
transition. The possibility is constrained by the laws of energy and
momentum conservation. Figure~\ref{fig1}(a) shows the allowed momentum-energy
regions in which phonons could excite an electron-hole pair and be
absorbed. We find that a phonon can only induce the intra-band transition
because the speed of sound is much smaller than the Fermi velocity
in real materials.

We first determine the phonon damping for a single Weyl node. For
a acoustic wave propagating along the magnetization axis ($z$-axis)
with $q_{z}<2k_{F}-\omega/v_{F}$, we obtain~\cite{supplementary}:
\begin{align}
\gamma_{\pm}^{D} & =\gamma_{0}\left(\frac{\left|q_{z}\right|}{k_{F}}+\frac{1}{4}\frac{q_{z}^{2}\left|q_{z}\right|}{k_{F}^{3}}\right)+\mathcal{O}\left(\frac{c_{T}}{v_{F}}\right),\label{gammaD}\\
\gamma_{\pm}^{A} & =\pm\eta\gamma_{0}\frac{q_{z}\left|q_{z}\right|}{k_{F}^{2}}+2\gamma_{0}\frac{q_{z}^{2}}{k_{F}^{2}}\frac{k_{F}}{k_{0}}\frac{c_{T}}{v_{F}}+\mathcal{O}\left(\frac{c_{T}^{2}}{v_{F}^{2}}\right),\label{gammaA}
\end{align}
where $\gamma_{0}\equiv(3/16)\pi\hbar\left(n_{e}/\rho_{I}\right)k_{0}^{2}$
with $n_{e}=k_{F}^{3}/3\pi^{2}$\textcolor{red}{{} }being the electron
number density and $\rho_{I}$ the ion mass density, the subscript
``$\pm$'' denotes the chirality of the Weyl node, and $\eta=\pm1$
depending on whether the Fermi level of electrons is located in the
upper band ($+1$) or the lower band ($-1$). $\gamma^{A}$ and $\gamma^{D}$
are both expanded in powers of $\omega$, which is substituted by
$\omega_{q_{z},L/R}\approx c_{T}q_{z}$. We ignore the small splitting
induced by the dissipationless viscosity because the contribution
is of the higher order ($c_{T}\ll v_{F}$, $\Delta c_{T}\ll c_{T}$).
For $q_{z}>2k_{F}-\omega/v_{F}$, the damping coefficients rapidly
decay to zero.

\begin{figure}
\includegraphics[width=0.48\columnwidth]{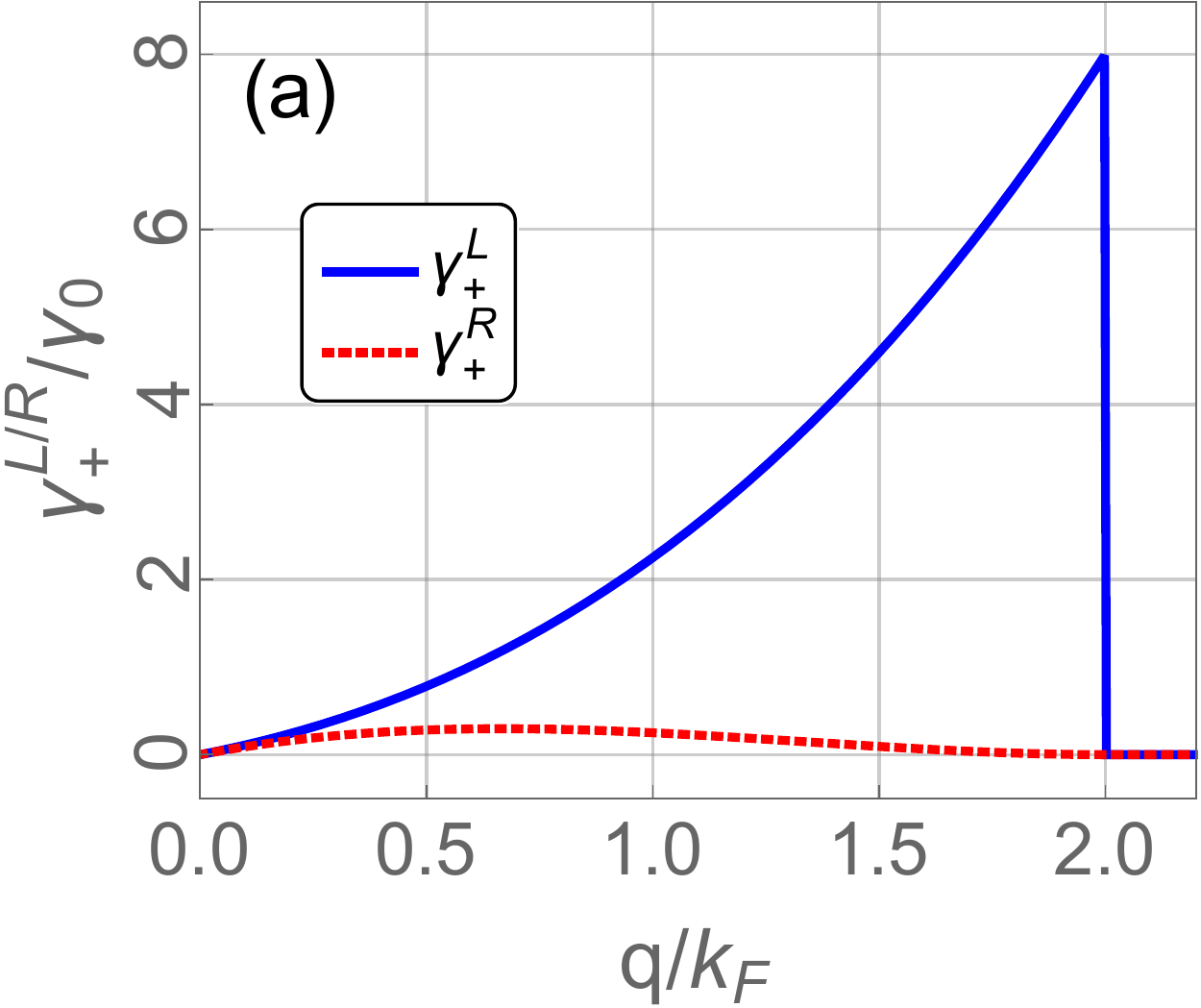}\quad{}\includegraphics[width=0.48\columnwidth]{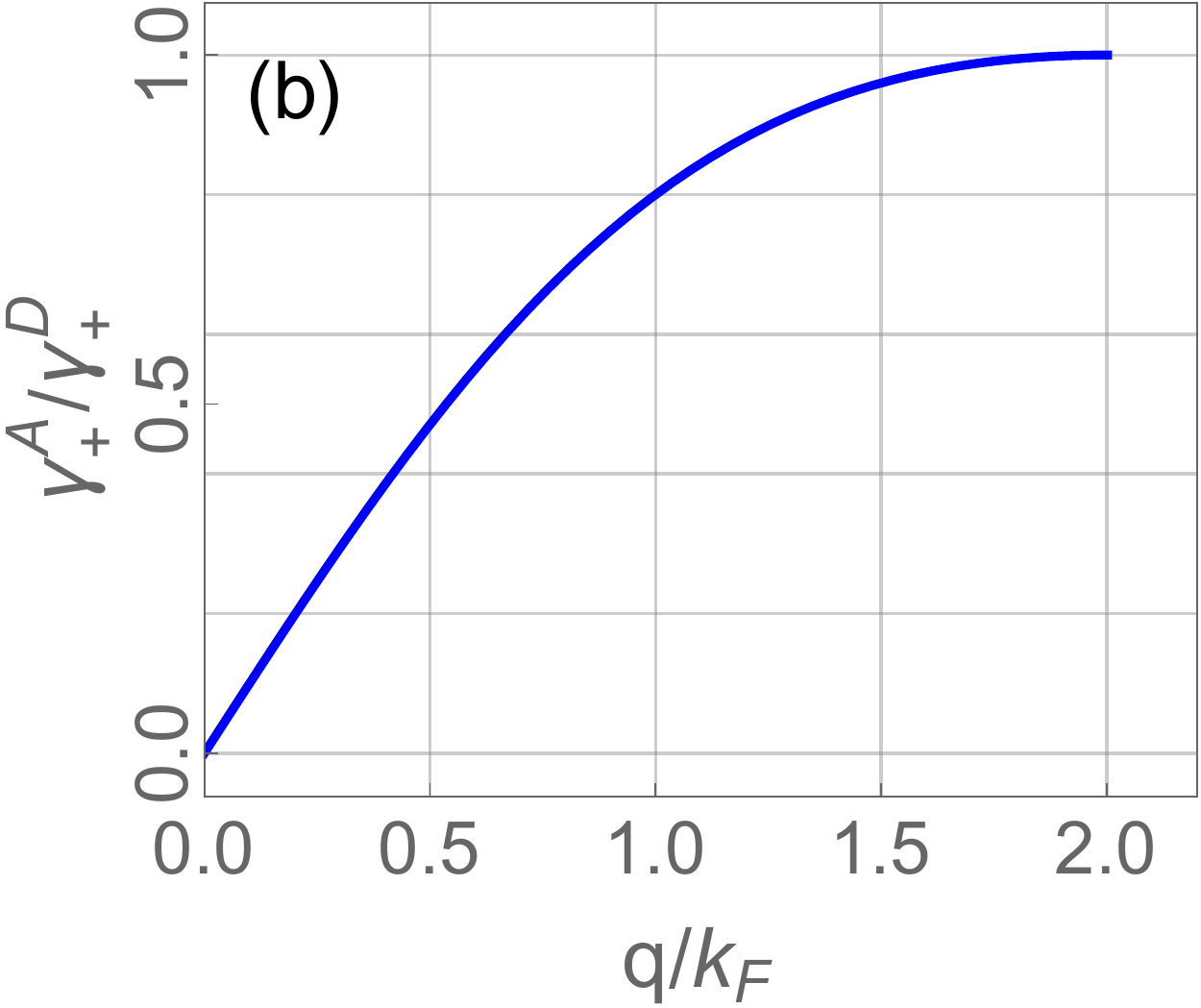}

\includegraphics[width=0.48\columnwidth]{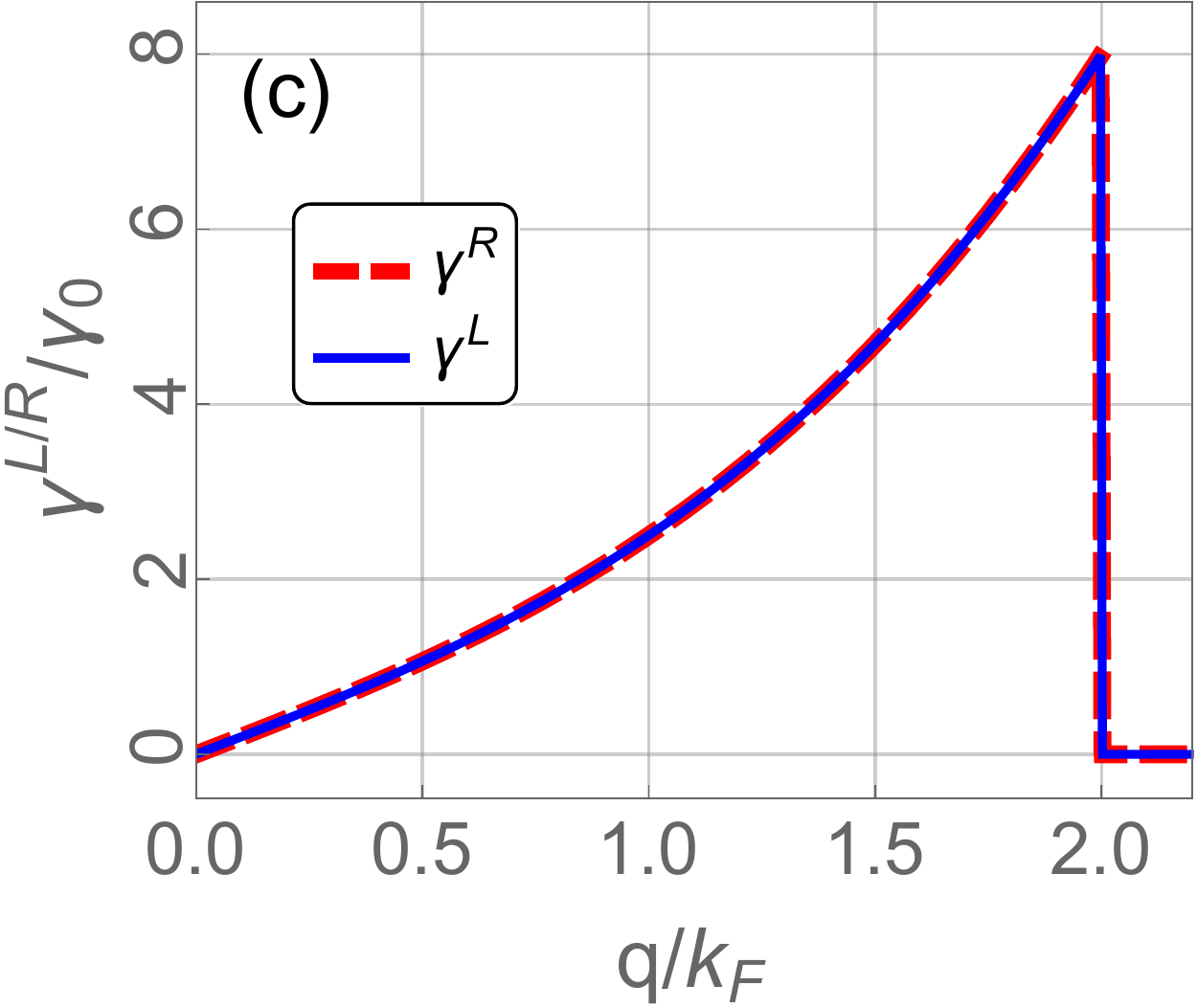}\quad{}\includegraphics[width=0.48\columnwidth]{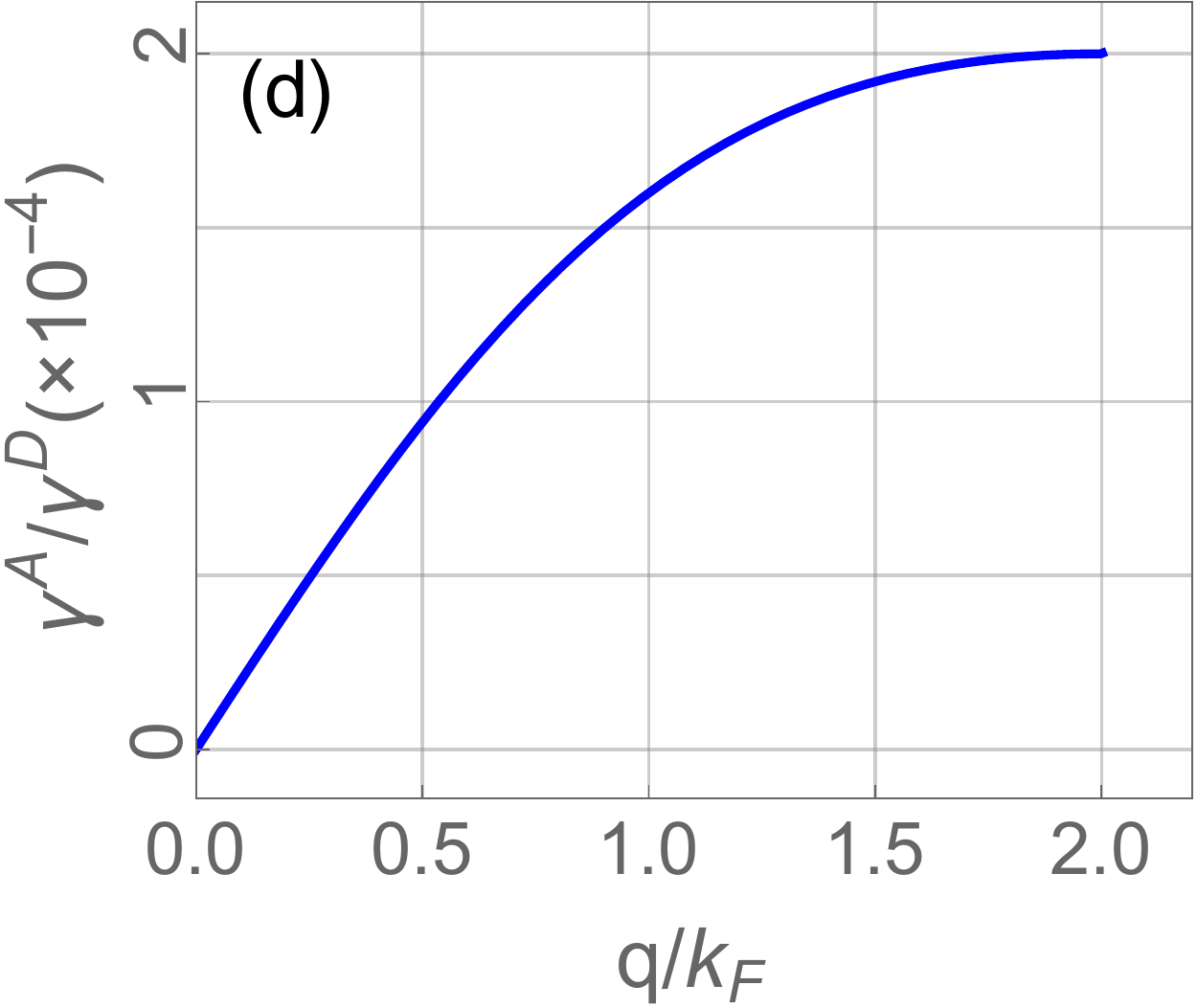}

\caption{\label{fig2} Damping coefficients for LCP ($\gamma^{L}$) and RCP
($\gamma^{R}$) phonon modes and the strengths of the circular dichroism
$\gamma^{A}/\gamma^{D}$ versus the phonon wave number $q_{z}$ for
a single left-handed ($+$) Weyl node (a, b) and a pair of Weyl nodes
(c, d) in a $\mathcal{I}$ symmetric system with $k_{0}=10k_{F}$,
$c_{T}=10^{-3}v_{F}$. }
\end{figure}

The phonon damping coefficient of a single left-handed ($+$) Weyl
node is shown in Fig.~\ref{fig2}(a, b). We find that the damping
of the LCP phonons is much stronger than that of the RCP phonons.
The relative circular dichroism $\gamma_{+}^{A}/\gamma_{+}^{D}$,
shown in Fig.~\ref{fig2}(b), is an increasing function of $q_{z}$,
vanishing at $q_{z}=0$ and reaching a maximum of $100\%$ at $q_{z}=2k_{F}$.
Because the energy of an electron is approximately conserved when
absorbing a phonon, the initial and final states of the electron must
be close to the Fermi surface. As a result, the selectivity is derived
from the spin texture on the Fermi surface. It is maximized (minimized)
when the spin directions of the initial and final states are anti-parallel
(parallel) to each other, as shown by Fig.~\ref{fig1}(b). For a
single Weyl node of the right-handed chirality ($-$), the leading
contribution in $\gamma^{A}$ will flip a sign. In this case, the
damping of the RCP phonons will be stronger.

Because Weyl nodes always appear in pairs, we sum contributions from
a pair of Weyl nodes of opposite chirality. In systems without breaking
$\mathcal{I}$, the two Weyl nodes related by the symmetry are identical
except the chirality. From Eq.~(\ref{gammaA}), we find the total
circular dichroism $\gamma^{A}=\gamma_{+}^{A}+\gamma_{-}^{A}=4\gamma_{0}(q_{z}/k_{F})^{2}(k_{F}/k_{0})(c_{T}/v_{F})$,
with the leading contribution in $\gamma_{\pm}^{A}$ cancelled. It
results in a strongly suppressed total circular dichroism. It is non-vanishing
but tiny, as shown in Fig.~\ref{fig2} (c-d).

\begin{figure}
\includegraphics[width=1\columnwidth]{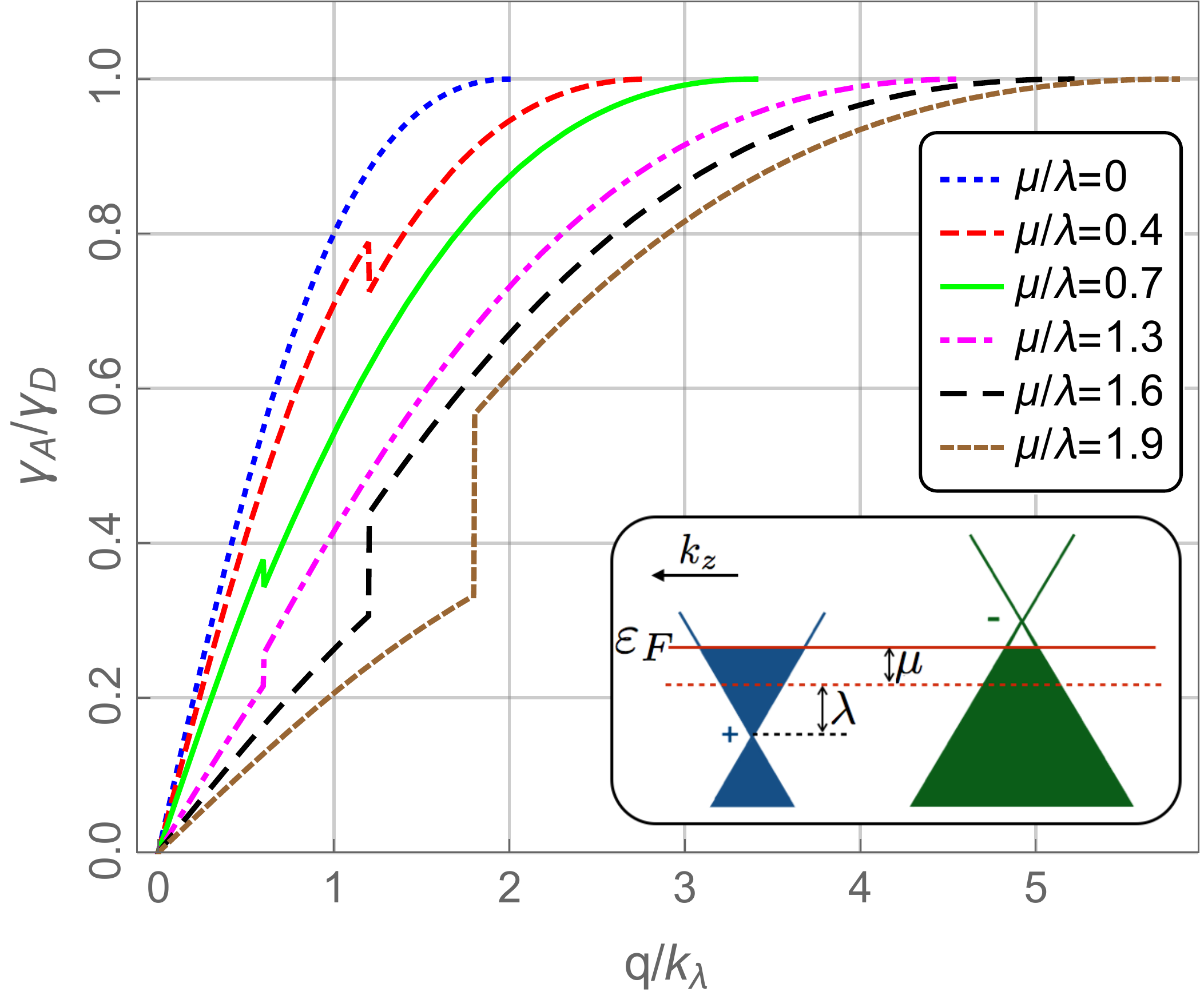}

\caption{\label{fig4} Relations of $\gamma_{A}/\gamma_{D}$ versus $q_{z}$
for different chemical potential $\mu$ when $\mathcal{I}$ is broken.
$k_{\lambda}\equiv\lambda/\left(\hbar v_{F}\right)$. The discontinuity
for the case of non-zero doping is due to the different sizes of the
two Fermi spheres, and occurs at the maximal phonon absorption wave-number
of the Weyl node with the smaller $k_{F}$.}
\end{figure}

The circular phonon dichroism can be greatly enhanced by breaking
$\mathcal{I}$ ($\lambda\neq0$), which shifts the two Weyl nodes
to different energies $\varepsilon^{\pm}=\mp\lambda$, as shown in
the inset of Fig.~\ref{fig4}. The effect is clearly manifested for
the case of zero doping ($\mu=0$). In this case, the Fermi level
is crossing the upper band ($\eta=1$) of one of the two Weyl nodes
and the lower band ($\eta=-1$) of the other one, giving rise to an
electron-like Fermi sphere and a hole-like Fermi sphere of the same
sizes. As a result, from Eq.~(\ref{gammaA}), the leading contributions
in $\gamma_{\pm}^{A}$ are not canceling each other out but adding
up, and the total circular dichroism $\gamma^{A}$ will be the double
of that from a single node. For the more general cases of non-zero
doping $(\mu\ne0$), the sizes of the two Fermi spheres will be different
with $k_{F}^{\pm}=(1/\hbar v_{F})\left|\lambda\pm\mu\right|$, and
the Fermi spheres could become both electron-like (hole-like) when
$|\mu/\lambda|\geq1$. Figure~\ref{fig4} shows the total relative
circular dichroism for different positions of the Fermi level. We
find that the giant circular dichroism is gradually suppressed when
the system moves away from the zero doping.\textcolor{blue}{{} }The
relative circular dichroism for $\mu\ne0$ shows a sharp drop or rise
at $q_{z}\approx2k_{F}^{-}$ due to the vanishing of the contribution
of the ``$-$'' Weyl node. In essence, the breaking of $\mathcal{I}$
unearths the giant circular dichroism inherent to each of the Weyl
nodes.

\noindent \textit{Experimental realization.\textemdash }The circular
phonon dichroism can be detected by measuring the difference of the
attenuation of the LCP/RCP acoustical waves in ultrasonic experiments.
We can estimate the damping coefficients for a magnetic Weyl semimetal.
We choose parameters $\rho_{I}\sim5\ \text{g}/\text{\ensuremath{\text{cm}^{3}}}$,
$n_{e}\sim10^{17}\ \text{cm}^{-3}$, $k_{0}\sim0.1\ \text{Å}^{-1}$
and $c_{T}\sim10^{3}\ \text{m}/\text{s}$, which are typical values
for known magnetic WSMs and Dirac semimetals~\cite{Hirschberger,Liu,Liu1,Jeon,Xiong}.
We obtain $\gamma_{0}\sim1.24\times10^{3}\ \text{s}^{-1}$. At $q=2k_{F}$
when the circular dichroism maximizes, the difference of the attenuation
for the LCP/RCP waves is the order of $\Delta\alpha=16\gamma_{0}/c_{T}\sim20\,\text{m}^{-1}\approx0.1\ \mathrm{dB/mm}$.
An alternative detection is to inject a linearly polarized TA wave
along a high symmetry magnetization axis. Due to the different phonon
damping rates of the LCP and RCP modes, the linearly polarized wave
will gradually become elliptically polarized and finally circularly
polarized. The presence of the dissipationless viscosity will also
induce a rotation of the major axis of the ellipse of the polarization.
The latter is the counterpart of the acoustic Faraday rotation observed
in magnetic insulators~\cite{Sytcheva}. The effect may find practical
applications. For instance, WSMs breaking both $\mathcal{T}$ and
$\mathcal{I}$ can be used as acoustic polarizers, which generate
circularly polarized acoustic waves from linearly polarized sources. 

\noindent \textit{Summary and Discussion.\textemdash }In summary,
we have shown that in magnetic metals with strong SOC, both the dissipationless
viscosity and the dissipative viscosity emerge. In particular, the
dissipative viscosity will induce the circular phonon dichroism. It
\textcolor{black}{offers a new degree of manipulation of phonons,
}\textcolor{black}{\emph{i.e.}}\textcolor{black}{, the control of
the phonon polarization.} We further show that WSMs breaking both
$\mathcal{T}$ and $\mathcal{I}$ exhibit giant circular phonon dichroism,
which provides a new characterization to these topologically nontrivial
materials and may find practical applications. 

Our study of the circular phonon dichroism in Weyl semimetals is limited
to the long-wave acoustic phonons. We note that the same effect also
presents for short-wave acoustic phonons as well as optical phonons.
The general formalism we have developed provides a solid foundation
and a unified approach for the calculation of the effect in real materials.
We further note that similar effects had been predicted in different
contexts in literatures. For instance, the Zeeman splitting of optical
phonons in multiferroic materials predicted in Ref.~\cite{Dynamical Multiferroicity}
is actually a manifestation of the dissipationless viscosity in the
optical branch of phonons. 
\begin{acknowledgments}
This work is supported by National Basic Research Program of China
(973 Program) Grant No. 2015CB921101 and National Science Foundation
of China Grant No. 11325416. 
\end{acknowledgments}

\widetext\clearpage{}

\setcounter{equation}{0}

\setcounter{figure}{0} 

\setcounter{table}{0}

\setcounter{page}{1}

\makeatletter

\global\long\def\theequation{S\arabic{equation}}

\global\long\def\thefigure{S\arabic{figure}}

\section*{Supplementary Materials for ``Circular Phonon Dichroism in Weyl
Semimetals''}

\section{Dissipationless viscosity in an insulator }

In the main text, we generalize the definition of the dissipationless
viscosity. Here, we show that $G_{\alpha\beta}$ defined in the main
text is exactly the effective magnetic field for phonons defined in
Ref.~\cite{Qin} for an insulator. We note that in the total Hamiltonian
$\hat{H}$ of the electron system, only $\hat{V}_{ei}$ depends on
the ion displacement $\bm{u}_{i}$. As a result, we can rewrite the
response function Eq.~(\ref{eq:response}) as: 
\begin{equation}
\mathcal{G}_{\alpha\beta}^{R}\left(\boldsymbol{R}_{n}-\boldsymbol{R}_{n'},t-t'\right)=-\frac{i}{\hbar}\theta\left(t-t'\right)\left\langle \left[\frac{\partial\hat{H}}{\partial u_{n}^{\alpha}}\left(t\right),\frac{\partial\hat{H}}{\partial u_{n'}^{\beta}}\left(t'\right)\right]\right\rangle _{0}.\label{definition}
\end{equation}
$\mathcal{G}_{\alpha\beta}^{R}$ can be expressed in the Lehmann representation
at zero temperature: 
\begin{align}
\mathcal{G}_{\alpha\beta}^{R}\left(\boldsymbol{R}_{n}-\boldsymbol{R}_{n'},\omega\right)= & \sum_{m\neq0}\frac{\left\langle \Phi_{0}\right|\frac{\partial\hat{H}}{\partial u_{n}^{\alpha}}\left|\Phi_{m}\right\rangle \left\langle \Phi_{m}\right|\frac{\partial\hat{H}}{\partial u_{n'}^{\beta}}\left|\Phi_{0}\right\rangle }{\hbar\omega+\epsilon_{0}-\epsilon_{m}+i\varepsilon}-\frac{\left\langle \Phi_{0}\right|\frac{\partial\hat{H}}{\partial u_{n'}^{\beta}}\left|\Phi_{m}\right\rangle \left\langle \Phi_{m}\right|\frac{\partial\hat{H}}{\partial u_{n}^{\alpha}}\left|\Phi_{0}\right\rangle }{\hbar\omega-\epsilon_{0}+\epsilon_{m}+i\varepsilon},\label{Lehmann}
\end{align}
where $\left|\Phi_{m}\right\rangle $ denotes an eigenstate with an
energy $\epsilon_{m}$ , and $\left|\Phi_{0}\right\rangle $ the ground
state. For an insulator, there is a gap separating the ground state
from the excited states, $\epsilon_{m}-\epsilon_{0}>0$ ($m\ne0$).
In this case, the dissipative (anti-hermitian) part of response function
vanishes, and the infinitesimal factor $i\varepsilon$ can be neglected
in Eq.~(\ref{Lehmann}). Following the definition shown in the main
text, we have: 
\begin{equation}
G_{\alpha\beta}\left(\boldsymbol{R}_{n}-\boldsymbol{R}_{n'}\right)=-i\frac{d\mathcal{G}_{\alpha\beta}^{R}\left(\boldsymbol{R}_{n}-\boldsymbol{R}_{n'},\omega\right)}{d\omega}\mid_{\omega\rightarrow0}.\label{insulatorG}
\end{equation}
We obtain: 
\begin{equation}
G_{\alpha\beta}\left(\boldsymbol{R}_{n}-\boldsymbol{R}_{n'}\right)=i\hbar\sum_{m\neq0}\frac{\left\langle \Phi_{0}\right|\frac{\partial H}{\partial u_{n}^{\alpha}}\left|\Phi_{m}\right\rangle \left\langle \Phi_{m}\right|\frac{\partial H}{\partial u_{n'}^{\beta}}\left|\Phi_{0}\right\rangle -\left\langle \Phi_{0}\right|\frac{\partial H}{\partial u_{n'}^{\beta}}\left|\Phi_{m}\right\rangle \left\langle \Phi_{m}\right|\frac{\partial H}{\partial u_{n}^{\alpha}}\left|\Phi_{0}\right\rangle }{\left(\epsilon_{0}-\epsilon_{m}\right)^{2}}.\label{formG}
\end{equation}
Note that the adiabatic wave function $\Phi_{m}$ depends on parameters
$\{\boldsymbol{u}\}$ (or $\{\boldsymbol{R}\}$).

To simplify, we employ identities: 
\begin{align}
\left\langle \left.\frac{\partial\Phi_{l}}{\partial u_{n}^{\alpha}}\right|\Phi_{m}\right\rangle  & =\frac{\left\langle \Phi_{l}\left|\frac{\partial\hat{H}}{\partial u_{n}^{\alpha}}\right|\psi_{m}\right\rangle }{\epsilon_{l}-\epsilon_{m}},\\
\left\langle \Phi_{l}\left|\frac{\partial\Phi_{m}}{\partial u_{n}^{\alpha}}\right.\right\rangle  & =\frac{\left\langle \Phi_{l}\left|\frac{\partial\hat{H}}{\partial u_{n}^{\alpha}}\right|\Phi_{m}\right\rangle }{\epsilon_{m}-\epsilon_{l}},
\end{align}
for $l\ne m$. We have: 
\begin{equation}
G_{\alpha\beta}\left(\boldsymbol{R}_{n}-\boldsymbol{R}_{n'}\right)=i\hbar\sum_{m}\left(\left\langle \left.\frac{\partial\Phi_{0}}{\partial u_{n}^{\alpha}}\right|\Phi_{m}\right\rangle \left\langle \Phi_{m}\left|\frac{\partial\Phi_{0}}{\partial u_{n'}^{\beta}}\right.\right\rangle -\left\langle \left.\frac{\partial\Phi_{0}}{\partial u_{n'}^{\beta}}\right|\Phi_{m}\right\rangle \left\langle \Phi_{m}\left|\frac{\partial}{\partial u_{n}^{\alpha}}\Phi_{0}\right.\right\rangle \right).\label{Gmid}
\end{equation}
Noting $\sum_{m}\left|\Phi_{m}\right\rangle \left\langle \Phi_{m}\right|=1$,
we obtain: 
\[
G_{\alpha\beta}\left(\boldsymbol{R}_{n}-\boldsymbol{R}_{n'}\right)=2\hbar\mbox{Im}\left\langle \left.\frac{\partial\Phi_{0}}{\partial u_{n'}^{\beta}}\right|\frac{\partial\Phi_{0}}{\partial u_{n}^{\alpha}}\right\rangle .
\]
This is exactly the effective magnetic field for phonons defined in
Ref.~\cite{Qin}.

\section{Viscosity coefficients of a non-interacting system\label{sec:Non-interacting}}

In this section, we derive the viscosity coefficients for a non-interacting
system.

For a non-interacting system, Eq.~(\ref{definition}) can be written
as: 
\begin{equation}
\mathcal{G}_{\alpha\beta}^{R}\left(\boldsymbol{q},\omega\right)=\frac{1}{M}\sum_{\boldsymbol{k}\boldsymbol{k}'ss'}\left\langle \psi_{\boldsymbol{k}s}\left|\frac{\partial\hat{v}}{\partial u_{-\boldsymbol{q}}^{\alpha}}\right|\psi_{\boldsymbol{k}'s'}\right\rangle \left\langle \psi_{\boldsymbol{k}'s'}\left|\frac{\partial\hat{v}}{\partial u_{\boldsymbol{q}}^{\beta}}\right|\psi_{\boldsymbol{k}s}\right\rangle \frac{f_{\boldsymbol{k}s}-f_{\boldsymbol{k}'s'}}{\epsilon_{\boldsymbol{k}s}-\epsilon_{\boldsymbol{k}'s'}+\hbar\omega+i\varepsilon},\label{Correlation_Hq}
\end{equation}
where $\hat{v}(\bm{r})=\sum_{\bm{R}}v_{ei}(\bm{r}-\bm{R})$ is the
single-particle interacting energy between an electron and all ions,
$\psi_{\boldsymbol{k}s}$ is the Bloch wave function at $\bm{k}$
and band $s$ with $\psi_{\boldsymbol{k}s}\left(\boldsymbol{r}\right)=N^{-1/2}e^{i\boldsymbol{k}\cdot\boldsymbol{r}}u_{{\boldsymbol{k}s}}\left(\boldsymbol{r}\right)$,
$\epsilon_{\boldsymbol{k}s}$ is the electron dispersion, and $\partial\hat{v}/\partial u_{\boldsymbol{q}}^{\alpha}$
is the Fourier transform of $\partial\hat{v}/\partial u_{n}^{\alpha}$:
\begin{equation}
\frac{\partial\hat{v}}{\partial u_{\boldsymbol{q}}^{\alpha}}=\frac{1}{\sqrt{N}}\sum_{n}\left(\frac{\partial\hat{v}}{\partial u_{n}^{\alpha}}\right)_{\{\boldsymbol{u}_{n}\}=0}e^{i\boldsymbol{q}\cdot\boldsymbol{R}_{n}^{0}}\equiv\frac{1}{2\sqrt{N}}\{\hat{T}_{\bm{q},\alpha},\,e^{i\bm{q}\cdot\bm{r}}\},\label{eq:Tq}
\end{equation}
where $\{\hat{T}_{\bm{q}},\,e^{i\bm{q}\cdot\bm{r}}\}$ denotes the
anti-commutator of the two operators.

By substituting Eq.~(\ref{eq:Tq}) into Eq.~(\ref{Correlation_Hq}),
we obtain: 
\begin{equation}
\mathcal{G}_{\alpha\beta}\left(\boldsymbol{q},\omega\right)=\frac{1}{MN}\sum_{\boldsymbol{k}ss'}\left\langle u_{\boldsymbol{k}s}\left|\hat{T}_{\bm{k},\bm{k}+\boldsymbol{q}}^{\alpha}\right|u_{\boldsymbol{k}+\boldsymbol{q}s'}\right\rangle \left\langle u_{\boldsymbol{k}+\boldsymbol{q}s'}\left|\hat{T}_{\bm{k}+\bm{q},\boldsymbol{k}}^{\beta}\right|u_{\boldsymbol{k}s}\right\rangle \frac{f_{\boldsymbol{k}s}-f_{\boldsymbol{k}+\boldsymbol{q}s'}}{\epsilon_{\boldsymbol{k}s}-\epsilon_{\boldsymbol{k}+\boldsymbol{q}s'}+\hbar\omega+i\varepsilon},\label{Correlation_Tq}
\end{equation}
where 
\begin{equation}
\hat{T}_{\bm{k},\bm{k}^{\prime}}^{\alpha}=\frac{1}{2}\left(e^{-i\bm{k}\cdot\bm{r}}\hat{T}_{\bm{k}-\bm{k}^{\prime},\alpha}e^{i\bm{k}\cdot\bm{r}}+e^{-i\bm{k}^{\prime}\cdot\bm{r}}\hat{T}_{\bm{k}-\bm{k}^{\prime},\alpha}e^{i\bm{k}^{\prime}\cdot\bm{r}}\right),\label{eq:Tkq}
\end{equation}

Using the definitions of the viscosity coefficients: 
\begin{align}
G_{\alpha\beta}\left(\boldsymbol{q}\right) & =-\frac{i}{2}\lim_{\omega\rightarrow0}\frac{d}{d\omega}\left(\mathcal{G}_{\alpha\beta}^{R}+\mathcal{G}_{\alpha\beta}^{R*}\right),\label{AntiHermitianG}\\
\gamma_{\alpha\beta}\left(\boldsymbol{q},\omega\right) & =\frac{i}{4\omega}\left(\mathcal{G}_{\alpha\beta}^{R}-\mathcal{G}_{\alpha\beta}^{R*}\right),\label{HermitianGamma}
\end{align}
we obtain their explicit forms:
\begin{align}
G_{\alpha\beta} & =\frac{i\hbar}{MN}\sum_{ss'\boldsymbol{k}}\left(f_{\boldsymbol{k}s}-f_{\boldsymbol{k}+\boldsymbol{q}s'}\right)\frac{\left\langle \varphi_{\boldsymbol{k}s}\right|\mathnormal{\hat{T}_{\bm{k},\bm{k}+\boldsymbol{q}}^{\alpha}}\left|\varphi_{\boldsymbol{k}+\boldsymbol{q}s'}\right\rangle \left\langle \varphi_{\boldsymbol{k}+\boldsymbol{q}s'}\right|\hat{T}_{\bm{k}+\bm{q},\boldsymbol{k}}^{\beta}\left|\varphi_{\boldsymbol{k}s}\right\rangle }{\left(\epsilon_{\boldsymbol{k}s}-\epsilon_{\boldsymbol{k}+\boldsymbol{q}s'}\right)^{2}+\varepsilon^{2}},\label{eq:Galphabeta}\\
\gamma_{\alpha\beta} & =\frac{\pi}{2MN\omega}\sum_{ss'\boldsymbol{k}}\left(f_{\boldsymbol{k}s}-f_{\boldsymbol{k}+\boldsymbol{q}s'}\right)\delta\left(\epsilon_{\boldsymbol{k}s}-\epsilon_{\boldsymbol{k}+\boldsymbol{q}s'}+\hbar\omega\right)\left\langle \varphi_{\boldsymbol{k}s}\right|\hat{T}_{\bm{k},\bm{k}+\boldsymbol{q}}^{\alpha}\left|\varphi_{\boldsymbol{k}+\boldsymbol{q}s'}\right\rangle \left\langle \varphi_{\boldsymbol{k}+\boldsymbol{q}s'}\right|\hat{T}_{\bm{k}+\bm{q},\boldsymbol{k}}^{\beta}\left|\varphi_{\boldsymbol{k}s}\right\rangle .\label{eq:Gamma}
\end{align}

\section{Kramer-Kronig relation}

Because both the dissipationless and the dissipative viscosities are
derived from the retarded response function $\mathcal{G}_{\alpha\beta}^{R}\left(\omega\right)$,
they are inherently related. To see that, we make use of the Kramers-Kronig
relation of $\mathcal{G}_{\alpha\beta}^{R}\left(\omega\right)$\cite{Vignale}:

\begin{equation}
\mathcal{G}_{\alpha\beta}^{R}\left(\bm{q},\omega\right)=-i\mathcal{P}\int_{-\infty}^{+\infty}\frac{\mathcal{G}_{\alpha\beta}^{R}\left(\bm{q},\nu\right)}{\nu-\omega}\frac{d\nu}{\pi},\label{KK correlation}
\end{equation}
where $\mathcal{P}$ denotes the Cauchy principal value. Substituting
Eq.~(\ref{KK correlation}) into Eq.~(\ref{AntiHermitianG}) and
making use Eq.~(\ref{HermitianGamma}), we obtain

\begin{align}
G_{\alpha\beta}(\bm{q}) & =-\frac{i}{2}\lim_{\omega\rightarrow0}\frac{d}{d\omega}\left[-i\mathcal{P}\int_{-\infty}^{+\infty}\frac{\mathcal{G}_{\alpha\beta}^{R}\left(\bm{q},\nu\right)-\mathcal{G}_{\beta\alpha}^{R*}\left(\bm{q},\nu\right)}{\nu-\omega}\frac{d\nu}{\pi}\right]\nonumber \\
 & =-2\lim_{\omega\rightarrow0}\frac{d}{d\omega}\left[-i\mathcal{P}\int_{-\infty}^{+\infty}\frac{\nu\gamma_{\alpha\beta}\left(\bm{q},\nu\right)}{\nu-\omega}\frac{d\nu}{\pi}\right]\nonumber \\
 & =2i\mathcal{P}\int_{-\infty}^{+\infty}\frac{\gamma_{\alpha\beta}\left(\bm{q},\nu\right)}{\nu}\frac{d\nu}{\pi}.\label{KK for G}
\end{align}

The positive frequency and the negative frequency components of the
$\gamma$ coefficients are related \cite{Vignale}:

\begin{equation}
\gamma_{\alpha\beta}\left(\boldsymbol{q},-\omega\right)=\gamma_{\beta\alpha}\left(-\boldsymbol{q},\omega\right).\label{eq:gammaSym}
\end{equation}
We obtain Eq.~(\ref{eq:KK}) by substituting Eq.~(\ref{eq:gammaSym})
into (\ref{KK for G}).

\section{the dissipationless viscosity of a deformation-dependent Haldane
Model}

We study a Haldane Model~\cite{Haldane} which is coupled to the
lattice deformation by
\begin{equation}
H=\epsilon\left(k\right)+d_{i}\left(k\right)e_{a}^{i}\sigma^{a},
\end{equation}
where $\epsilon\left(\boldsymbol{k}\right)=2t_{2}\cos\phi\sum_{i}\cos\left(\boldsymbol{k}\cdot\boldsymbol{b}_{i}\right)$,
$d_{x}\left(\boldsymbol{k}\right)=t_{1}\sum_{i}\cos\left(\boldsymbol{k}\cdot\boldsymbol{a}_{i}\right)$,
$d_{y}\left(\boldsymbol{k}\right)=t_{1}\sum_{i}\sin\left(\boldsymbol{k}\cdot\boldsymbol{a}_{i}\right)$,
$d_{z}\left(\boldsymbol{k}\right)=\epsilon_{0}-2t_{2}\sin\phi\left(\sum_{i}\sin\left(\boldsymbol{k}\cdot\boldsymbol{b}_{i}\right)\right)$,
$i=1,2,3$, $\boldsymbol{a}_{i}$ and $\boldsymbol{b}_{i}$ are vectors
connecting nearest and second- nearest sites, $t_{1}$ and $t_{2}\exp(i\phi)$
are the hopping constants of the nearest and next-nearest neighbors,
$\epsilon_{0}$ is the on-site energy ($+\epsilon_{0}$ and $-\epsilon_{0}$
for two sub-lattices). The lattice constant is $a$. For $\left|t_{2}/t_{1}\right|<1/3$,
the Chern number is non-zero when $\left|\epsilon_{0}/t_{2}\right|<3\sqrt{3}\left|\sin\phi\right|$.

\begin{figure}
\includegraphics[scale=0.5]{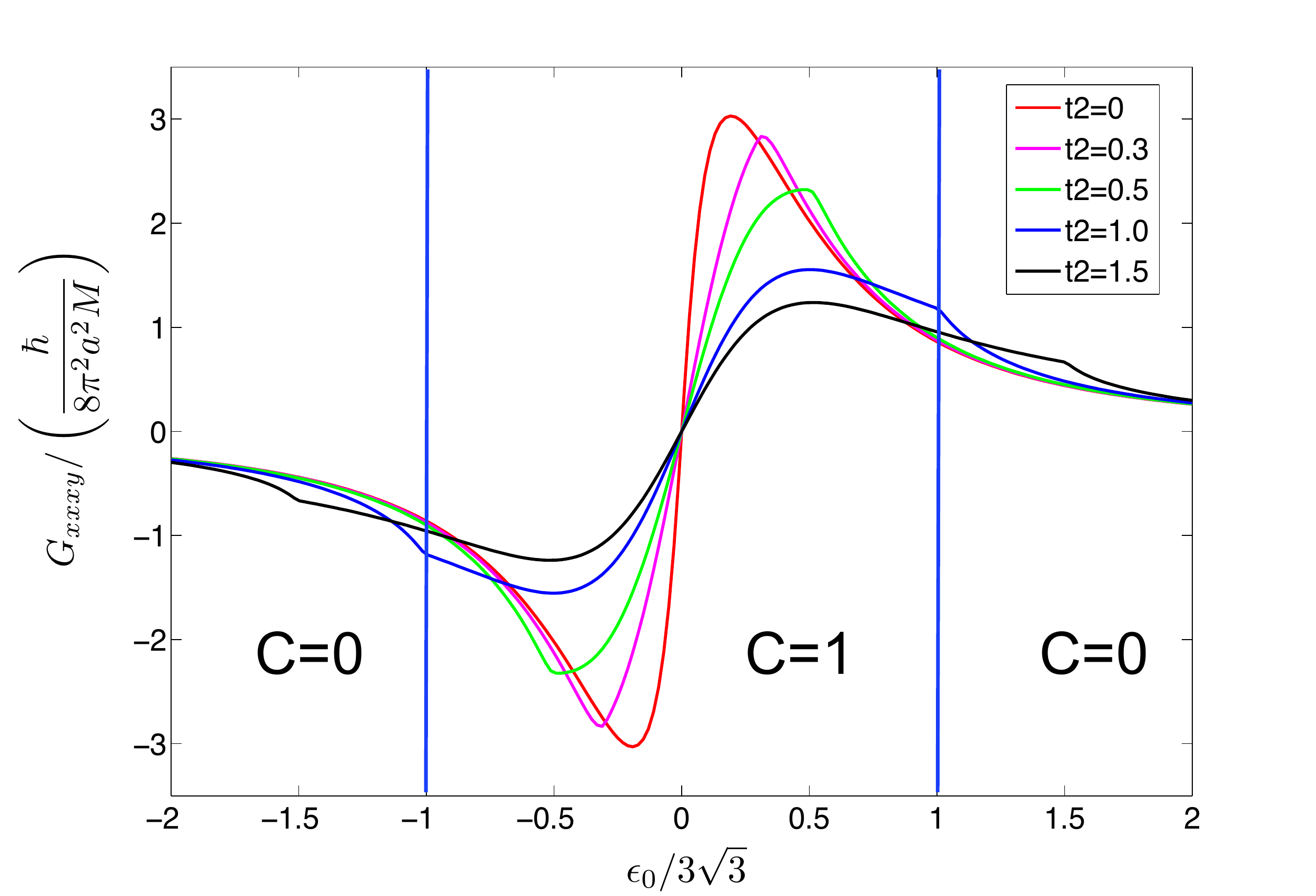}

\caption{\label{fig:The-dissipationless-viscosity}The dissipationless viscosity
of a Haldane model coupling with lattice deformation. }
\end{figure}

We set $\phi=\pi/2$, and calculate the dissipationless viscosity.
The relation of the dissipationless viscosity and parameter $\epsilon_{0}$
for different $t_{2}$ is shown in Fig.~\ref{fig:The-dissipationless-viscosity}.
It shows that the dissipationless viscosity is nonzero even when the
Chern number is zero, although there appears a kink at the point of
the transition from a topologically nontrivial regime to a trivial
one. 

In Fig.~\ref{fig:The-dissipationless-viscosity}, $G_{xxxy}$ is
defined as
\begin{equation}
G_{xxxy}=\frac{1}{2}\lim_{\bm{q}\rightarrow0}\frac{\partial^{2}}{\partial q_{x}^{2}}G_{xy},
\end{equation}
and can be calculated by:

\begin{equation}
G_{xxxy}=\frac{1}{MV}\frac{d}{d\omega}\int_{-\infty}^{+\infty}dte^{i\omega t}\left[-\frac{i}{\hbar}\theta\left(t-t'\right)\right]\left\langle \left[\frac{\partial H}{\partial e_{x}^{x}}\left(t\right),\frac{\partial H}{\partial e_{y}^{x}}\right]\right\rangle _{0}.\label{eq:viscosity xxxy}
\end{equation}

\section{Circular phonon dichroism of a single Weyl node\label{CPDWN}}

We first consider a left-handed Weyl node. The eigenstates of the
low energy effective Hamiltonian Eq.~(\ref{Weyl Hamiltonian}) in
the absence of the lattice deformation can be written as $\psi_{\eta,\boldsymbol{k}}\left(\boldsymbol{r}\right)=V^{-\frac{1}{2}}e^{i\boldsymbol{k}\cdot\boldsymbol{r}}\chi_{\eta,\tilde{\boldsymbol{k}}}$
with $\eta=\pm$ denoting the upper and the lower band, respectively,
$\tilde{\boldsymbol{k}}\equiv\boldsymbol{k}-k_{0}\hat{z}$ , and

\begin{align}
\chi_{+,\tilde{\boldsymbol{k}}} & =\left(\begin{array}{c}
\cos\frac{\theta_{\tilde{\bm{k}}}}{2}\\
\sin\frac{\theta_{\tilde{\bm{k}}}}{2}e^{i\varphi_{\tilde{\bm{k}}}}
\end{array}\right),\\
\chi_{-,\tilde{\boldsymbol{k}}} & =\left(\begin{array}{c}
-\sin\frac{\theta_{\tilde{\bm{k}}}}{2}e^{-i\varphi_{\tilde{\bm{k}}}}\\
\cos\frac{\theta_{\tilde{\bm{k}}}}{2}
\end{array}\right),
\end{align}
where $\theta_{\tilde{\bm{k}}}$ and $\varphi_{\tilde{\bm{k}}}$ are
the polar and azimuthal angles of the vector $\tilde{\boldsymbol{k}}$.
The corresponding eigen-energies are $\varepsilon_{\boldsymbol{k}\eta}=\eta\hbar v_{F}\left|\tilde{\boldsymbol{k}}\right|$.

Using the definition Eq.~(\ref{eq:Tkq}) and the effective Hamiltonian
Eq.~(\ref{Weyl Hamiltonian}), we obtain: 
\begin{equation}
\hat{T}_{\bm{k},\boldsymbol{k}^{\prime}}^{\alpha}=\frac{i\hbar v_{F}}{2}\left(\bm{k}-\bm{k}^{\prime}\right)\cdot\left(\bm{k}+\bm{k}^{\prime}\right)\hat{\sigma}^{\alpha}.\label{eq:Tkqw}
\end{equation}
Here, we have set $N=V$ when applying Eq.~(\ref{eq:Tkq}). It is
a convenient choice for a continuous system defined by Eq.~(\ref{Weyl Hamiltonian}).

We have shown in the main text that phonons can only induce the intra-band
transition in a Weyl node. Only the term with $s=s'=\eta$ in Eq.
(7) is contributing to $\gamma_{\alpha\beta}$. By substituting Eq.~(\ref{eq:Tkqw})
into Eq.~(\ref{eq:Gamma}), we obtain: 
\begin{equation}
\gamma_{\alpha\beta}^{+}\left(\boldsymbol{q},\tilde{\omega}\right)=\frac{\pi\hbar}{2\tilde{\omega}M}\int\frac{dk^{3}}{\left(2\pi\right)^{3}}F_{\eta}^{\alpha\beta}\left(\tilde{\boldsymbol{k}},\tilde{\boldsymbol{k}}'\right)\left[\boldsymbol{q}\cdot\left(\boldsymbol{k}+\frac{\boldsymbol{q}}{2}\right)\right]^{2}\left(f_{\boldsymbol{k}\eta}-f_{\boldsymbol{k}+\boldsymbol{q}\eta}\right)\delta\left(\tilde{\omega}+\eta\left|\tilde{\boldsymbol{k}}\right|-\eta\left|\tilde{\boldsymbol{k}}+\boldsymbol{q}\right|\right),
\end{equation}
where the superscript ``$+$'' denotes the left-handed chirality
of Weyl node, $\tilde{\omega}=\omega/v_{F}$, and 
\begin{equation}
F_{\eta}^{\alpha\beta}\left(\tilde{\boldsymbol{k}},\tilde{\boldsymbol{k}}'\right)=\left\langle \chi_{\eta,\tilde{\boldsymbol{k}}}\right|\hat{\sigma}^{\alpha}\left|\chi_{\eta,\tilde{\boldsymbol{k}}+\boldsymbol{q}}\right\rangle \left\langle \chi_{\eta,\tilde{\boldsymbol{k}}+\boldsymbol{q}}\right|\hat{\sigma}^{\beta}\left|\chi_{\eta,\tilde{\boldsymbol{k}}}\right\rangle .
\end{equation}

It is straightforward to evaluate the matrix elements and obtain :
\begin{align}
F_{\eta}^{xy}\left(\boldsymbol{k};\boldsymbol{k}'\right) & =\frac{i\eta\left(k_{z}k'-kk_{z}'\right)+k_{x}k_{y}'+k_{y}k_{x}'}{2kk'},\\
F_{\eta}^{xx}\left(\boldsymbol{k},\boldsymbol{k}'\right) & =\frac{kk'-k_{z}k_{z}'-k_{y}k_{y}'+k_{x}k_{x}'}{2kk'}.
\end{align}

Completing the integral, we obtain the full expression for the damping
coefficients for phonons with $\bm{q}=(0,0,q_{z})$, 
\begin{align}
\gamma_{xy}^{+} & =-i\gamma_{0}\mbox{sign}\left(q_{z}\right)\frac{\left(q_{z}^{2}-\tilde{\omega}^{2}\right)}{k_{F}^{2}}\left(\eta+\frac{2k_{F}\tilde{\omega}}{k_{0}q_{z}}+\eta\frac{k_{F}^{2}\tilde{\omega}^{2}}{k_{0}^{2}q_{z}^{2}}+\frac{\tilde{\omega}^{3}}{6k_{F}k_{0}q_{z}}+\eta\frac{\tilde{\omega}^{4}}{4k_{0}^{2}q_{z}^{2}}\right),\label{gammaxy}\\
\gamma_{xx}^{+} & =\gamma_{0}\mbox{sign}\left(q_{z}\right)\frac{\left(q_{z}^{2}-\tilde{\omega}^{2}\right)}{k_{F}^{2}}\left[\left(\frac{k_{F}}{q_{z}}+\frac{q_{z}}{4k_{F}}\right)+\eta\left(\frac{2k_{F}^{2}}{k_{0}q_{z}^{2}}+\frac{1}{2k_{0}}\right)\tilde{\omega}+\left(\frac{k_{F}^{3}}{k_{0}^{2}q_{z}^{3}}+\frac{k_{F}}{4k_{0}^{2}q_{z}}+\frac{1}{12k_{F}q_{z}}\right)\tilde{\omega}^{2}+\eta\frac{\tilde{\omega}^{3}}{2k_{0}q_{z}^{2}}\right.\nonumber \\
 & \left.+\left(\frac{k_{F}}{2k_{0}^{2}q_{z}^{3}}+\frac{1}{48k_{F}k_{0}^{2}q_{z}}\right)\tilde{\omega}^{4}+\frac{\tilde{\omega}^{6}}{80k_{F}k_{0}^{2}q_{z}^{3}}\right],\label{gammaxx}
\end{align}
where $\gamma{}_{0}=\frac{3}{16}\pi\hbar\left(n_{e}/\rho_{I}\right)k_{0}^{2}$,
and the formulas are valid only when $|q_{z}|<2k_{F}-\tilde{\omega}$.
The damping coefficients vanishes when $|q_{z}|>2k_{F}+\tilde{\omega}$.
In the intermediate regime $2k_{F}-\tilde{\omega}<|q_{z}|<2k_{F}+\tilde{\omega}$,
the damping coefficients rapidly decay to zero.

Because the left- and right-handed Weyl nodes can be related to each
other by $\mathcal{I}$ , the damping coefficients of a right handed
Weyl node can be obtained by the relation $\gamma_{\alpha\beta}^{-}\left(\boldsymbol{q}\right)=\gamma_{\alpha\beta}^{+}\left(-\boldsymbol{q}\right)$.

For the long-wave TA phonon with the dispersion $\omega=c_{T}q_{z}$,
$\tilde{\omega}_{q_{z},L/R}=\left(c_{T}/v_{F}\right)q_{z}$ is a small
quantity since $c_{T}\ll v_{F}$. By keeping the leading order terms
in Eq.~(\ref{gammaxy}) and (\ref{gammaxx}), we obtain Eqs.~(\ref{gammaD},\ref{gammaA}).

\section{Elastic gauge field}

In Eq.~(\ref{Weyl Hamiltonian}), we only consider the minimum coupling
between electrons and the lattice deformation. An alternative way
to introduce the coupling is through an elastic gauge field. For the
elastic gauge field, we adopt the effective Hamiltonian obtained in
Ref.~\cite{Cortijo} : 
\begin{equation}
H=\boldsymbol{\sigma}\cdot\left(\hbar v_{F}\boldsymbol{k}_{\perp}\pm\boldsymbol{A}_{\perp}\right)\mp\sigma_{z}\left(\hbar v_{F}k_{z}\pm A_{z}\right),\label{elasticH}
\end{equation}
where we assume that the Fermi velocity is isotropic. The elastic
gauge fields are expressed as, in our notation: 
\begin{align}
A_{x}^{el} & =\beta b_{3}u_{zx},\\
A_{y}^{el} & =\beta b_{3}u_{zy},\\
A_{z}^{el}= & 2\beta b_{3}u_{zz}-\sum_{j}\beta ru_{jj},
\end{align}
where $u_{ij}$ is the strain tensor, and $\beta$ is the Grüneisen
parameter~\cite{Shapourian}. In tight-binding models, $\beta$ is
defined as $\beta=-\partial\ln t/\partial\ln a\simeq2$, where $t$
and $a$ are the hopping integral and the distance between two sites,
respectively.

Following the same procedure as shown in Sec.~\ref{CPDWN}, we obtain
the leading contributions of the elastic gauge field: 
\begin{align}
\gamma_{\pm}^{D} & =\gamma'_{0}\left(\frac{\left|q_{z}\right|}{k_{F}}+\frac{1}{4}\frac{q_{z}^{2}\left|q_{z}\right|}{k_{F}^{3}}\right)+\mathcal{O}\left(\frac{c_{T}^{2}}{v_{F}^{2}}\right),
\end{align}

\begin{align}
\gamma_{\pm}^{A} & =\pm\eta\gamma'_{0}\frac{q_{z}\left|q_{z}\right|}{k_{F}^{2}}+\mathcal{O}\left(\frac{c_{T}^{2}}{v_{F}^{2}}\right).
\end{align}
Compared with Eq.~(\ref{gammaD}) and (\ref{gammaA}), $\gamma_{0}$
is changed to $\gamma'_{0}=(\beta^{2}/4)\gamma_{0}$, scaled by a
factor $\beta^{2}/4$. Since $\beta\simeq2$, the elastic gauge field
yields essentially the same result as that from the minimum coupling.

\end{document}